# Adsorption at Nanoconfined Solid-Water Interfaces

Anastasia G. Ilgen*[1], Kevin Leung[1], Louise J. Criscenti[1], and Jeffery A. Greathouse[2]

1. Sandia National Laboratories, Geochemistry Department, PO Box 5800 Mailstop 0754, Albuquerque, NM 87185-0754, United States

2. Sandia National Laboratories, Nuclear Waste Disposal Research and Analysis, PO Box 5800 Mailstop 0779, Albuquerque, NM 87185-0779, United States

| Name | ORCID ID | Email |
|---|---|---|
| Anastasia G. Ilgen | 0000-0001-7876-9387 | agilgen@sandia.gov |
| Kevin Leung | 0000-0003-1397-3752 | kleung@sandia.gov |
| Louise J. Criscenti | 0000-0002-5212-7201 | ljcrisc@sandia.gov |
| Jeffery A. Greathouse | 0000-0002-4247-3362 | jagreat@sandia.gov |

*Corresponding author. E-mail agilgen@sandia.gov

## Abstract

Reactions at solid-water interfaces play a foundational role in water treatment systems, catalysis, chemical separations, and in predicting chemical fate and transport in the environment. Over the last century, experimental measurements and computational models have made tremendous progress in capturing reactions at solid surfaces. The interfacial reactivity of a solid surface, however, can change dramatically and unexpectedly when it is confined to the nanoscale. Nanoconfinement can arise in different geometries such as pores/cages (3-D confinement), channels (2-D confinement) and slits (1-D confinement). Therefore, measurements on unconfined surfaces, and molecular models parameterized based on these measurements, fail to capture chemical behaviors under nanoconfinement. This review evaluates recent experimental and theoretical advances, with a focus on adsorption at solid-water interfaces. We review how nanoconfinement alters the physico-chemical properties of water, and how the structure and dynamics of nanoconfined water dictate energetics, pathways, and products of adsorption in nanopores. The implications of these findings and future research directions are discussed.

# 1. Introduction

Adsorption at unconfined solid-water interfaces has been studied for decades, but recent breakthroughs in both experimental techniques and modeling have enabled a new generation of adsorption studies in more complex nanoconfined systems (~0.5-500 nm). This review evaluates how the energetics, pathways, and products of adsorption are driven by the structure and dynamics of nanoconfined systems and spans both environment- and technology-relevant examples (Figure 1). Research shows that the geometry of nanoconfined systems is important; however, due to the limited number of relevant studies, we included all pore geometries where adsorption was studied, including pores/cages (3-D confinement), channels (2-D confinement) and slits (1-D confinement), and we collectively refer to these systems as "nanopores."

Earth's surface has numerous nanoconfined solid-water interfaces, and reactivity of these nanopores define the fate and transport of chemical species (1). Natural nanopores include clay mineral interlayers, zeolite pores, nanoporous sedimentary rocks, and nano-cracks in soil/rock grains. Natural nanoporous zeolites and clay minerals are widely utilized in industry as reactive substrates or supports (2; 3). In new material discovery and separation sciences, synthetic and natural nanoporous architectures are at the forefront of critical applications including gas separations (4), water purification (5), critical material recovery (6), nanocatalysis (7), energy storage (8), and sensing (9) (Figure 1).

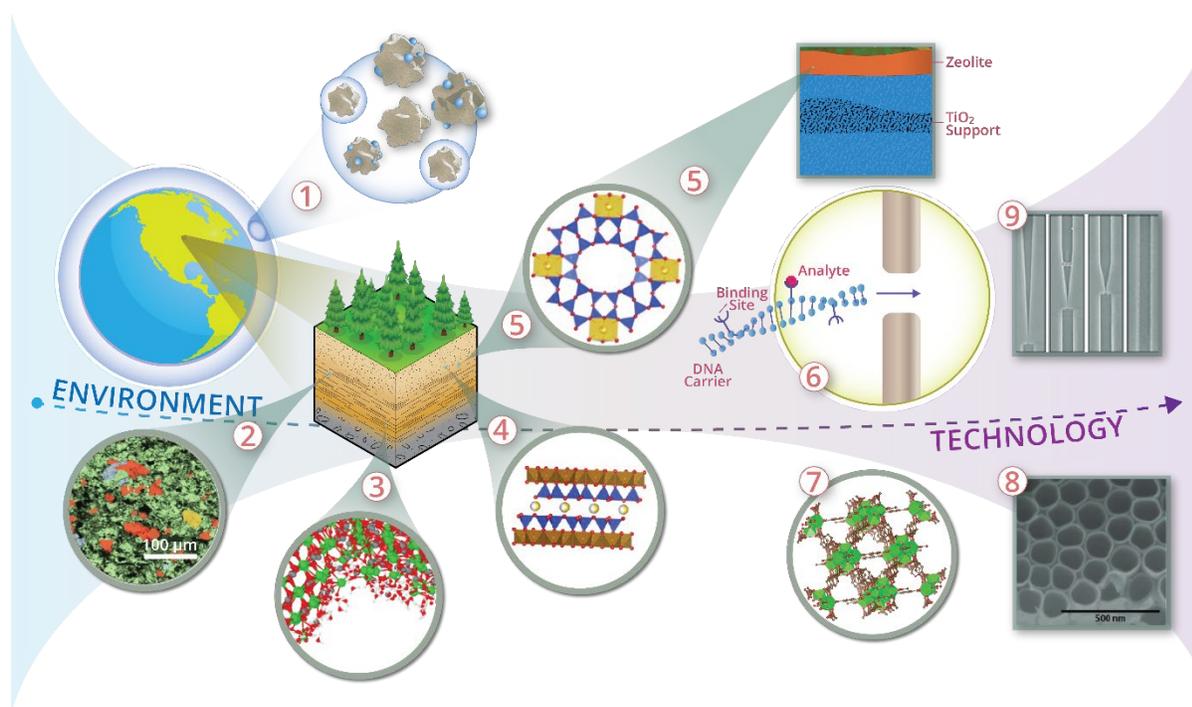

**Figure 1.** Nanoconfined solid-water interfaces are important in the fate and transport of elements in the environment and in technological applications. (1) atmospheric dust with nanopores and nm-scale thin water films; (2) nanopores in soil particles and sedimentary rocks; (3) nano-scale cracks in rocks; (4) nanochannels in clay minerals; (5) nanocages in natural zeolites, which are also used in industrial applications of catalysis, ion exchange, and in hierarchical membranes with nanoporous active layers, from Ref. (2); (6) nanopore sensing and chemical analysis (adapted from Albrecht Annu. Rev. Anal. Chem. 2019. 12:371–87); (7) Synthetic metal organic frameworks UiO-66; (8) nanoporous $TiO_2$ electrode in water splitting applications, from Ref. (10); (9) nanofluidic «lab-on-a-chip», from Ref. (11). Mineral structures obtained from American Mineralogist Crystal structure database and visualized with the Vesta 3.3.9 program.

The reason reactive nanopores are of great interest is that surfaces exhibit unexpected and often beneficial reactivities at the nanoscale, including lowering reaction barriers and stabilizing thermodynamically unstable phases (1); however, questions remain about the combined effects of nanopore dimensions and surface chemistry on interfacial reactivity. Specifically, for nanopores filled with $H_2O$ or aqueous solutions, nanoconfinement-driven changes to the structure and dynamics of $H_2O$ have been characterized and quantified (12-15), and yet, we cannot predict how nanoconfinement affects the free energies of solute-solute and solute-surface interactions, and therefore the adsorption pathways and products. Making these predictions is complicated, because the electrical double-layer (EDL) extending from the charged surface into the aqueous phase imposes structure on the chemical species in the interfacial regions (H-bonding and the preferred orientation of $H_2O$ dipoles; the mean distances between ions and surfaces) (16-21). In nanopores, overlapping EDLs from opposing surfaces increase the complexity of the interfacial reactions and the difficulty of both characterization and accurate predictions.

Furthermore, nanoconfined $H_2O$ has lower effective dielectric response (relative permittivity) in single-digit (<10 nm) nanopores compared to bulk $H_2O$ (22). This is expected, based on the studies of unconfined solid-water interfaces, where $H_2O$ structure and preferred orientation near solid surfaces leads to a decrease in the dielectric response (23). Additionally, this decrease has been reported for very large nanochannels (<1000 nm) that extend beyond the range of van der Waals (vdW) forces (24) or the Debye length $\kappa$ of most electrolytes (*i.e.* the inverse square-root of the salt concentration) (25). At room temperature, $\kappa$ varies widely between energy storage systems (~ 0.3 nm in 1 M monovalent electrolytes), surface waters (~10 nm in ~1 mM salt solutions), and "pure" water (1000 nm in $10^{-7}$ M $H^+$/$OH^-$ concentrations). Nanoconfined $H_2O$ has lower density, surface tension, and freezing point (13; 20; 26-39) explained by its distinct translational, rotational, and vibrational behavior (12; 14; 16). Furthermore, nanoconfined $H_2O$ can exhibit higher apparent viscosity in nanochannels (40-42), which affect solute and $H_2O$ diffusion through geological or membrane nanopores (see section 2.5).

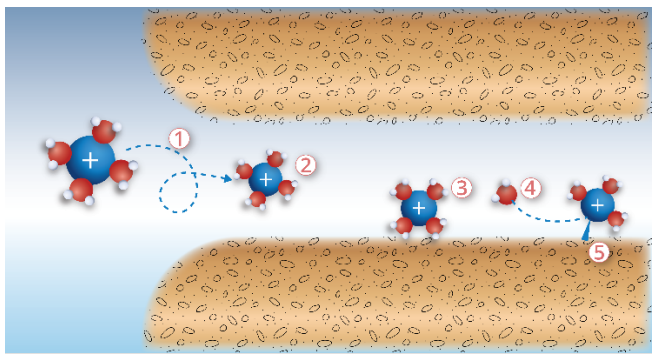

**Figure 2.** Elementary reaction steps and structures in nanopores that define adsorption: (1) Solvated ion enters a nanopore; (2) solvated ion diffuses through a nanopore; (3) solvated ion adsorbs on nanopore surface as outer-sphere complex (physisorption); (4) partial ion de-solvation inside nanopores; (5) partially de-solvated ion adsorbs as inner-sphere complex (chemisorption).

Here we summarize experimental and theoretical findings for adsorption in solid nanopores filled with aqueous solutions, where solvated aqueous species and surface sites react to produce outer-sphere (physisorption) or inner-sphere (chemisorption) surface complexes (Figure 2). We include aluminosilicates (zeolites, clay minerals, mica minerals) as well as templated oxides in this review, and exclude soft confining materials, such as reverse micelles. De-solvation of either an adsorbing species or a reactive surface site is required for the formation of an inner-sphere surface complex. Additionally, (de)protonation, and (de)hydroxylation reactions of either aqueous species or surface sites often accompany adsorption.

Therefore, to better understand how nanoconfinement affects adsorption, we first review the physico-chemical properties of nanoconfined $H_2O$, since they define ion diffusion into nanopores, and solvation free energies for the reactive surface sites and for dissolved/adsorbed species. In Section 2 we summarize classical theories of overlapping EDL structures and discuss why they are insufficient to describe the structure and energetics of nanoconfined systems. We then review the structures, dynamics, dielectric properties of $H_2O$ and $H_2O$ self-dissociation in nanoconfined systems — the processes which could control adsorption. In Section 3 we show that nanoconfinement changes the products (favoring inner-sphere complexation), pathways, and thermodynamics of adsorption. Finally, in Section 4 we outline future research needs for understanding and predicting chemical forces, molecular structures and interfacial reactivities in nanopores filled with aqueous solutions. Understanding these fundamental phenomena will accelerate material design for key applications, including the ion-selective capture of pollutants and recovery of critical elements , wastewater treatment, and desalination.

## 2. Emergent phenomena in nanopores

### 2.1 Forces in nanopores

The classical approach to describe the governing forces between charged surfaces in solution is based on Deryaguin-Landau-Verwey-Overbeek (DLVO) theory which states that the total potential energy $V_T$ of interaction between two infinite flat plates is given as:

$$V_T = V_R + V_A \qquad \text{(Eq. 1)}$$

where $V_R$ is the repulsive term due to electrostatic interaction between two identical surfaces and $V_A$ is the attractive vdW interaction between these surfaces. The Poisson-Boltzmann (PB) equations are used to describe $V_R$ and the Gouy-Chapman (GC) model expands on the PB equations to describe the distribution of counterions in a diffuse layer formed at a single charged surface but does not include the compact (Stern) layer where ions adsorb to the surface as outer-sphere or inner-sphere complexes. These classical models have simplifying assumptions (e.g., the use of point charge ions) that are periodically revisited. Ninham and co-workers (43) argue that the separation of electrostatic and quantum mechanical forces in Eq. (1) is thermodynamically incorrect. Nonetheless, DLVO theory is used as a starting point to understand charged surfaces in solution.

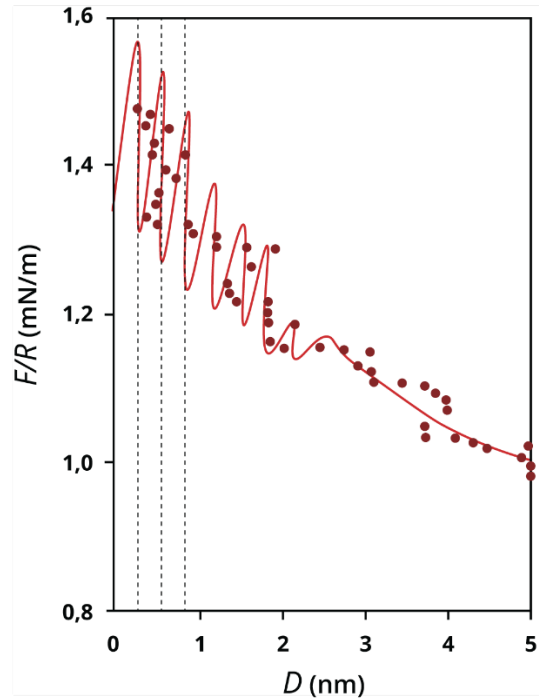

**Figure 3.** Single force curve for a mica-silica system in a $10^{-4}$ M NaCl at pH ~5.1. Oscillations have a period of ~0.25 nm, close to the diameter of $H_2O$ molecule, From Ref. (44).

Numerous studies show the failure of DLVO theory at small separations. Hunter (45) addressed this by adding a solvation term $V_S$ to Eq. (1) that accounts for the influence of a surface on adjacent water layers and decays exponentially at ~1 nm distance from the surface. In aqueous solutions, positive $V_S$ indicates the presence of hydration forces; negative, attractive values are believed to indicate hydrophobic interaction. $V_S$ includes both short range oscillatory forces that arise whenever water molecules form quasi-discrete layers between two surfaces and a monotonic force that decays with surface separation caused by surface-water interactions (44) (Figure 3). The oscillations only exist near atomically smooth surfaces; a surface roughness of a few Å can eliminate them.

To account for small separations, DLVO theory for mica-mica systems was refined by incorporating molecular details: strongly adsorbed hydration layers, crystal lattice anisotropy, and ionizable surface groups (46). The new model successfully predicts the electrolyte dependence of the experimental data for mica surfaces in contact with 0 to 100 mM NaCl (46). With the addition of water structuring to the model it predicts that the total force becomes attractive at ~3 nm distance (pH~7, 1mM NaCl) and fits the experimental data better than standard DLVO theory (47). Ion-specific effects on the stepwise decreases in pull-off force for mica sheets in solutions containing $Na^+$, $K^+$, $Cs^+$ or $H_3O^+$ at <2 nm distance were attributed to the collective dehydration of cations (48). The oscillatory component of $V_S$ is independent of salt concentration and pH but dependent on cation type: it is unaffected by strongly hydrated cations ($Li^+$, $Na^+$) but suppressed by weakly hydrated cations ($Rb^+$, $Cs^+$) (49). The monotonic hydration force decreases in strength with decreasing cation Gibbs free energy of solvation ($\Delta G_{solv}$), leading to a transition from an overall repulsive $V_T$ ($Li^+$, $Na^+$) to an attractive $V_T$ ($Rb^+$, $Cs^+$) that can be detected out to 1.5 nm (49). Similar studies with silica exhibit a net short-range repulsive $V_T$ force at <2 nm distances, caused by overlapping hydrated layers and H-bond formation between Si-OH surface groups and neighboring $H_2O$ molecules (49). In mixed mica-silica systems, $V_S$ again includes a monotonic repulsive force overlain by an oscillatory force (44).

There is no single explanation for the short-range hydration forces (non-DLVO forces) that are not included in classical DLVO theory. These forces are present between silica surfaces in pure water but only in finite salt concentrations between mica surfaces. In silica systems, the force is reduced by the presence of electrolytes: cations with lower $\Delta G_{solv}$ produce weaker repulsion force. The opposite is observed in mica systems (e.g., Ref. (49)), illustrating that the hydration forces are a function of both solution composition and surface chemistry.

Clay minerals, like smectite, exhibit overlapping EDLs in their interlayer spaces. In humid air or high-ionic-strength solutions, smectites can accommodate between 1 and 3 ordered $H_2O$ layers, with interlayer spacings that vary in discrete steps from 1.1 to 1.9 nm (section 2.5). In low-ionic-strength solutions, smectites can incorporate more water expanding up to ~14 nm, depending on electrolyte type and concentration, permanent layer charge, and temperature. However, DLVO theory fails to predict the ionic-strength threshold for clay particle aggregation and swelling pressure, and accurate behavior for highly charged surfaces and divalent cations (50).

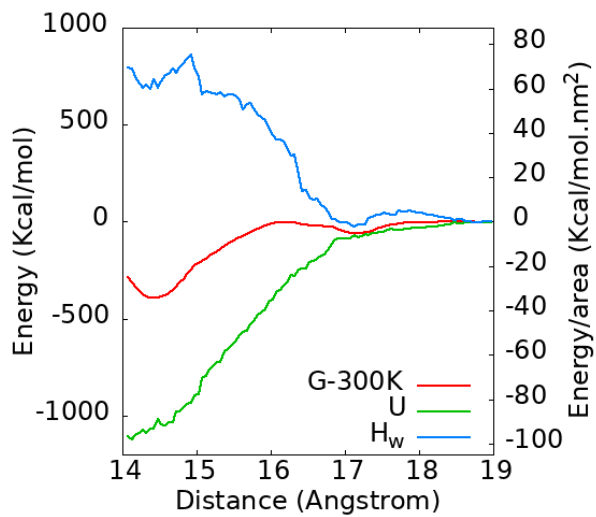

**Figure 4.** Thermodynamic decomposition of the PMF for oriented attachment between the basal surfaces of two gibbsite particles. The dominant energy components are U(r), the direct potential between the two particles, and Hw(r), the enthalpic contribution from water-particle and water-water interaction. Modified from Ref. (51).

Several models have been developed for clay minerals that combine the GC model with more details regarding the EDL overlap region. Goncalves *et al.* combined the triple-layer model with a truncated version of the GC model to explain the dependence of osmotic efficiency and swelling pressure of smectite as functions of mean pore size for interacting diffuse layers (52), and later included the combined effects of ion hydration and the electric field on water structure

(53). Gilbert *et al.* speculated that variations in interlayer spacings observed for smectite in the osmotic hydrate state arise from the action of vdW forces extending over multiple interlayer distances, including >5 smectite layers plus the intervening water (50). Classical molecular dynamics (CMD) simulations of smectite particles in liquid water and in 1M NaCl with $Na^+$, $K^+$, or $Ca^{2+}$ counterions confirm that a strong non-DLVO attraction arises at ~3 nm from ion-clay Coulomb interactions in the EDL (54). The breakdown of the DLVO model is most obvious for water films <1 nm, where disjoining pressures display strong repulsion or oscillations with characteristic length scales commensurate with the thickness of a water monolayer (0.3 nm). Short-range repulsion associated with the surface and ion hydration effects is suggested to create an energy barrier to dehydration below the two-layer hydrate. This theory is supported by the calculation of the potential of mean force (PMF) between two gibbsite ($Al(OH)_3$) surfaces to investigate the activation energy barriers from the two-layer hydrate to complete dehydration (Figure 4) (51). Ho and Criscenti found that the dominant contribution to the primary minimum is enthalpic and that the enthalpy associated with water-surface and water-water interactions is highly unfavorable to particle aggregation, similar to the short-range repulsive hydration forces in the mica and silica studies.

    Ninham and co-workers emphasizes that ionic dispersive forces, between ions and solvent, for ion-ion and ion-surface interactions are not explicit in the classical DLVO theory and that these missing ionic quantum fluctuation forces play a large role in specific ion effects and hydration. Using a semi-atomistic modeling approach, Dragulet *et al.* show that in clay-like interlayers, increasing confinement and surface charge densities promote ion-water structures that increasingly deviate from ions' bulk hydration shells, with ion valency and size determining these deviations (55). While it is increasingly clear that the nature of the solid surfaces, solution

compositions, ion size and valency, all contribute to the chemistry of nanoconfined solutions, we are far from establishing a comprehensive theory that predicts how these properties translate into apparent adsorption rates, pathways, and products in nanopores.

2.2 Hydrogen bonding structure and dynamics of nanoconfined water

In liquid water, $H_2O$ dipoles interact to form dynamic hydrogen-bonding (HB) networks that change with temperature, ionic composition, and proximity to a solid surface, even when the $H_2O$ does not form H-bonds with the surface (56; 57). Recent studies of unconfined $SiO_2$ surfaces with varying Si-OH surface site densities (i.e. varying abilities to form H-bonds with $H_2O$) show that the chemistry of the surface defines the HB structures in the interfacial regions (56). CMD simulations provided spatially resolved details of HB arrangements in nanopores (12; 57; 58). HB networks are distorted in hydrophilic $SiO_2$ pores within ~ 1 nm from the pore surface (i.e. fewer HBs have neighboring $H_2Os$, compared to bulk $H_2O$) (59), due to the competition between water-water and water-surface interactions. In hydrophobic carbon nanotubes, $H_2Os$ also form fewer HBs with each other, while $H_2O$ does not form HBs with the confining surfaces (60). CMD simulations often show that in the diffuse EDL region, HB networks can be indistinguishable from those in bulk $H_2O$ (12; 57; 58), although the re-arrangement dynamics (speed with which HB network responds during perturbation) is much slower in nanopores than in bulk $H_2O$ (61). This HB distortion in nanopores is akin to increasing temperature of bulk water, which may favor de-solvation and promote inner-sphere adsorption; however, no research has identified how the details of HB environments in nanopores control adsorption.

At present, it is impossible to obtain spatially resolved experimental information about HB networks in nanopores because signals come from the entire volume of the nanopore and its external surfaces, but averaged signals provide some useful characterization of HB networks. For example, vibrational spectra of the water OH stretching region shows three broad bands of $H_2Os$: (i) $H_2O$ molecules with zero HBs (free $H_2O$); (ii) $H_2O$ molecules which have between one and three HBs; and (iii) $H_2O$ molecules with four HBs. However, these observations cannot always exclude the contribution of resonant vibrational coupling at ~ 3200 $cm^{-1}$ (62) and fall short of the CMD-predicted five discrete $H_2O$ populations with zero, one, two, three, or four HB neighbors (28; 63; 64). Averaging between the distinct $H_2O$ populations (64; 65) creates apparent inconsistencies in the vibrational spectroscopy results: some indicate that the number of 4-coordinated $H_2O$ molecules (and the average number of HB) in hydrophilic $SiO_2$ pores increases with decreasing pore diameter, which disagrees with CMD predictions (28; 63); while others indicate a lower average number of HBs in nanoconfined $H_2O$, compared to bulk $H_2O$ (64). Future studies should focus on determining HB network characteristics (including water-water, and water-surface HBs) in nanopores as a function of pore dimensions, aqueous phase compositions, and surface chemistries.

## 2.3 Dielectric properties of $H_2O$ in nanoconfined domains

Cluster-based Density Functional Theory (DFT) with an implicit approach to describe $H_2O$ molecules has routinely been applied to model chemical reactions in liquids. However, the implicit solvation approach depends on the dielectric screening, which in turn depends on nanoconfinement. Dielectric response is also crucial for interpreting the predictions of periodic boundary condition *ab initio* molecular dynamics (AIMD) simulations of coupled protonation and ion desorption reactions at solid-water interfaces and in nanopores (66).

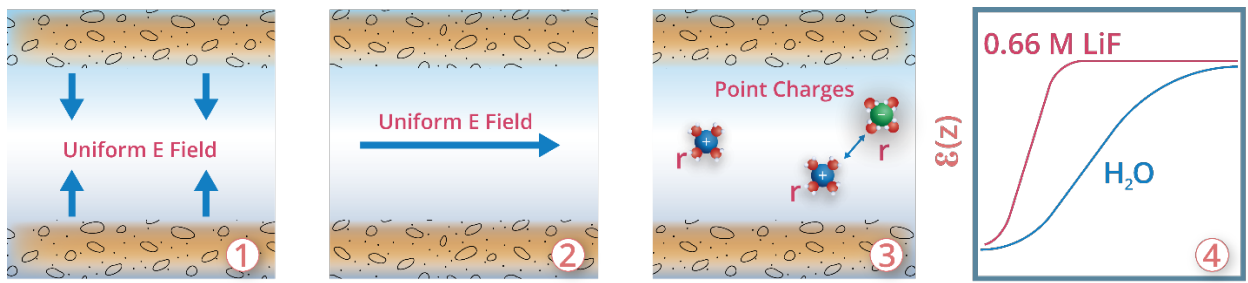

**Figure 5.** Three of the possible dielectric responses arising from different perturbations in a slit pore and are relevant to (1) $\varepsilon_\perp$; (2) $\varepsilon_\parallel$; and (c)$\varepsilon$ (**r**) and $\varepsilon$(**r**,**r'**). (1) is most frequently discussed in the literature, partly because it directly maps on to the capacitance of parallel plate capacitors and can be most readily measured. For ion-hydration, ion-ion, and ion-surface site interactions, (3) is the most relevant. (4) Dielectric profile adapted from experimental data from Ref. (67).

In a water-filled slit nanopore, the perpendicular and parallel dielectric tensor components ($\varepsilon_\perp$ and $\varepsilon_\parallel$) with respect to the nanopore surfaces can be differentiated (Figure 5). $\varepsilon_\parallel$ is typically predicted to be close to the bulk $H_2O$ value (~78), while $\varepsilon_\perp$, the only component which can be measured in a laboratory, was found to decrease to 2 at solid-water interfaces (67) and in sub-nanometer confined 2-D slit pores (22). To our knowledge, there are only three nanoconfined systems, the muscovite mica surface (67), boron nitride (22), and silica nanochannels (24), for which $\varepsilon_\perp$ were measured. Recent advances questioned the role of $\varepsilon_\perp$ in ion-ion interactions (68; 69). In slit pores, short range interactions between two ions at the same distance from the pore surface were predicted to depend on the geometric mean of $\varepsilon_\perp$ and $\varepsilon_\parallel$, and exhibit less nanoconfinement effect than if they were governed exclusively by $\varepsilon_\perp$ (69).

Calculated or measured dielectric profiles $\varepsilon_\perp(z)$ reveal spatially resolved information. Ballenegger and Hansen (70) derived a rigorous linear-response formula which was successfully applied for $H_2O$ confined between biological membranes, in boron nitride nanopores (71), and in other confined media, evaluated using CMD simulation. It appears to yield sharp peaks when applied to the interior of crystalline regions, suggesting that further development is needed for solid-water interfaces associated with crystalline surfaces with permanent charges.

The above discussion focuses on nanoconfined $H_2O$, which behaves like a classical dielectric material: the electric field is attenuated by a factor of $1/\varepsilon_\perp(z)$ at distances z from the surface (72). When ions are present at isolated surfaces, the dielectric response is qualitatively different. Electric fields are completely screened beyond a few Debye lengths, and (for electrochemical capacitor applications) the capacitance becomes an interfacial rather than a bulk property (73). To our knowledge, the calculated $\varepsilon_\perp(z)$ (74; 75) have yet to be demonstrated to exhibit the experimental Debye-Hückel form, shown in the 4$^{th}$ panel in Figure 5 (From Ref. (67)). Atomic simulations instead focused on the strongly nanoconfined regime, where the EDLs of two solid surfaces overlap, which may be most pertinent for environmental and nanocatalytic reactions, as well as permeation through membrane nanopores. A plethora of dielectric-related findings associated with nanoconfined chemical species remain to be reconciled. For example, $\varepsilon_\perp(z)$ were predicted to increase, and $\varepsilon_\parallel(z)$ decrease, nonlinearly in graphene slit pores as a function of increasing NaCl concentrations up to 3 M (74; 75), which conflicts with the expectation that higher salt concentrations lead to more charge screening and shorter extent of EDL (72; 73). The low dielectric constant of $H_2O$ directly at an isolated surface was suggested to encourage ion-pairing (68), which in turn affects dielectric properties. Specific ion effects on chemical reactivity, e.g., the *pKa* of silanol groups on $SiO_2$ surfaces, were measured (76) and predicted (77), suggesting that $\varepsilon_\perp(z)$ is also surface-chemistry-dependent. Future research should provide a cohesive framework to reconcile these observations, and should translate $\varepsilon_\perp(z)$ calculated in the presence of ions into chemical reactivity in nanopores and compare them with direct CMD predictions of $\varepsilon_\perp(z)$ (76; 77). It is however possible that certain ions need to be explicitly included in the reaction zone to depict specific ion effects. Finally, the above discussion focuses on slit pores. Because the derivations of dielectric profiles in slit (1-D confinement) pores and cylindrical (2-D confinement) pores are somewhat different, the role of geometries of nanoconfined systems represent a significant future research opportunity.

2.4 Self-dissociation of water in nanopores

Self-dissociation of $H_2O$ is the most fundamental aqueous reaction, yet it is unclear how nanoconfinement may affect it (78). When considering adsorption reactions, it is important to understand the acid-base properties of nanoconfined $H_2O$, because they will define aqueous speciation in nanopores, and therefore the charges of adsorbing species . The self-dissociation of $H_2O$ includes two explicit steps: breaking the H—OH covalent bond, and the diffusion and separation of $OH^-$ and $H_3O^+$ species. From the thermodynamic perspective, breaking the H—OH covalent bond has the highest energy barrier, and $H^+$ diffusion *via* the Grotthuss mechanism is nearly barrierless (79). These steps are associated with the HB network re-organization to accommodate newly created charged species. The energy required for H—OH bond breaking is dependent on the strength of the HB environment around a given $H_2O$ molecule, which depends on the number of HB neighbors (section 2.2).

At present, the studies on $H_2O$ self-dissociation in nanopores are contradictory. For example, AIMD simulations of mackinawite (FeS) nanopores, predict increased $H_2O$ self-dissociation. Researchers propose that the equilibrium constant $K_w$ increases from ~$10^{-14}$ to ~$10^{-12}$ (79) due to the nanoconfinement of $H_2O$ itself, and that water-surface interactions can be ignored (79). In contrast, the first-principles calculations by Di Pino *et al*. (80) show that $H_2O$ nanoconfinement in clusters/droplets has no effect on $H_2O$ self-dissociation for clusters that consist of >6 $H_2O$ molecules and conclude that surface-water interactions define self-dissociation of $H_2O$ in nanopores. An experimental study of single $H_2O$ molecules H-bonded within $C_{60}$ cages shows decreased ability to self-dissociate (81). AIMD simulations by Liu *et al.* predict that $H_2O$ self-dissociation in clay mineral interlayers is defined by the surface charge, and to a lesser degree by nanoconfinement (82). Qualitatively similar trends are predicted inside carbon nanotubes (83), while the role of nanoconfinement on proton behaviors in zeolites and resulting acid strengths is described as an "open debate" (84). To understand acid-base chemistry in nanopores, and how it shapes adsorption reactions, research is needed to resolve individual contributions from nanoconfinement (dimensions) vs. surface chemistry to the overall $H_2O$ self-dissociation equilibrium.

## 2.5 Diffusion of nanoconfined $H_2O$ and solutes in nanopores

Diffusion of aqueous species and water inside nanopores defines adsorption on nanopore surfaces. This diffusion is influenced by water-surface interactions ($H_2O$ flows "fast" in hydrophobic nanopores, and "slow" in hydrophilic nanopores) (85) and can be directly related to macroscopic observables (e.g. ion permeation through membranes or permeability and tracer diffusion in subsurface reservoirs). Nanoscale experimental measurements are enabled by quasielastic neutron scattering (QENS) (86; 87) and nuclear magnetic resonance (NMR) spectroscopy (88; 89). The average diffusion coefficients of CMD simulations can be directly compared with experimental results (90; 91), and the diffusion of $H_2O$ and solutes can also be assessed relative to the pore surface (90; 92-96).

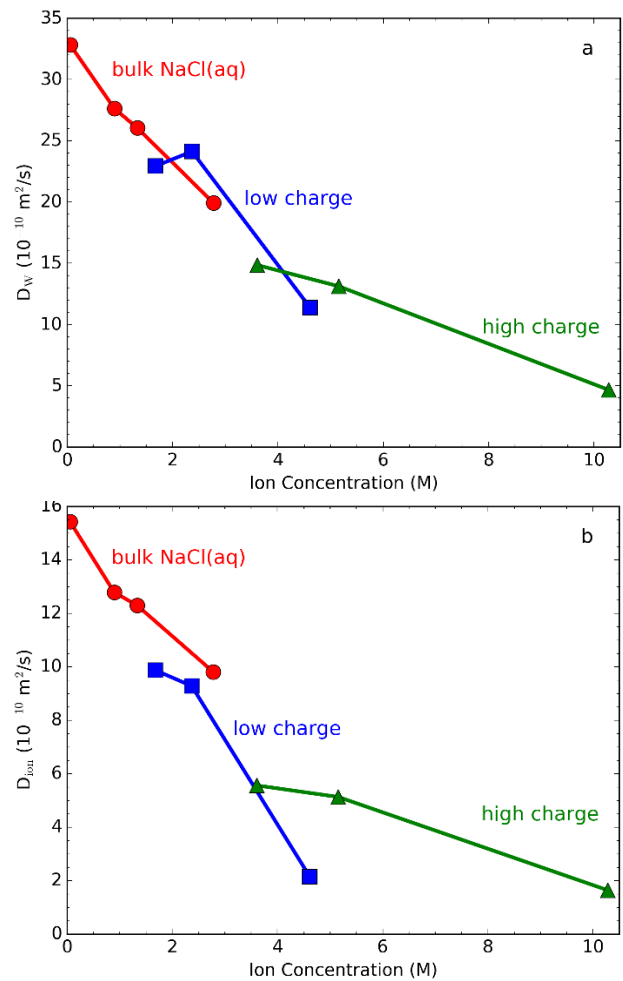

**Figure 6.** Results from CMD simulations of hydrated Na-montmorillonite interlayers (blue and green) and bulk NaCl solutions (red) at 300 K showing (a) water and (b) sodium ion diffusion coefficients as a function of ion concentration and clay layer charge (97). Pore size increases as ion concentration decreases in the Na-montmorillonite models. Images reprinted with permission from the American Chemical Society.

Experimental and simulation studies of hydrated smectite clay minerals show that $H_2O$ mobility increases as pore size (water content) increases. Macroscopic ion diffusion data have been combined with QENS to estimate the tortuosity of fluid diffusion in both clay interlayers and the pores between clay particles (98). Complex impedance spectroscopy has also been used to measure $H_2O$ diffusion in clay interlayers (99). In addition to pore size, lack of particle orientation (anisotropy) reduces $H_2O$ mobility in hydrated clay samples (88). Efforts have been made to connect fluid mobility in idealized clay interlayers with macroscopic clay aggregates which also include interparticle pores that are larger than clay interlayers (100; 101). Underwood and Bourg (101) show that $H_2O$ diffusion behavior approaches that of bulk clay interlayers as the clay particles become more ordered.

The decrease in $H_2O$ mobility in clays at lower water content is attributed to increasing ion concentration: as water content (and layer spacing) decrease $H_2O$ and ion mobility are dominated by ion-water interactions (Figure 6) (97). As cation concentration increases (pore size decreases), the fraction of solvating $H_2Os$ increases, which limits $H_2O$ mobility. The same trend is seen in corresponding bulk NaCl solutions. Similarly, $H_2O$ and cation diffusion decreases in clay interlayers filled with hydrophilic polymers due to strong water H-bonding with the intercalated polymer (15), and $H_2O$ diffusion in 1 nm silica nanopores with high surface charge is significantly reduced due to the presence of adsorbed counterions (92).

Aluminosilicate nanotubes (imogolite) provide a platform to investigate $H_2O$ diffusion in cylindrical nanopores in the presence of hydrophilic (hydroxyl terminated, IMO-OH) and hydrophobic (methyl terminated, IMO-$CH_3$) surfaces. Neutron scattering experiments show significant nanoconfinement effects on water diffusion in 1.5 nm imogolite pores regardless of surface chemistry, with greater mobility in hydrophobic IMO-$CH_3$ (85). At low water content, $H_2O$ diffusion in hydrophilic IMO-OH is reduced since the primary transport mechanism involves jump events between surface H-bonding sites. Even at full $H_2O$ saturation, the $H_2O$ diffusion coefficient in IMO-OH is approximately half that of bulk water (85). Simulations of $H_2O$ in imogolite with diameters between 1.0 – 1.5 nm show both fast surface diffusion and 1-D diffusion in the center of the pores that increases with pore size (102; 103). Simulations have also shown that the curvature of the outer $Al_2O_3$ surface prevents in-plane H-bonds, which affects water properties compared with the planar counterpart (1-D confinement in a slit pore) (104).

QENS measurements on templated $SiO_2$ MCM-41 with ~2 nm pores show similar $H_2O$ diffusion behavior in both hydrophilic (unmodified) and hydrophobic (modified with $CH_3$ surface groups) samples: diffusion slower than bulk water at temperatures below 300 K, but approximately the same as bulk water at 300 K (105). Similarly, QENS results for $SiO_2$ MCM-41 analogs with SiOH, AlOH, and ZrOH surface terminations show that $H_2O$ diffusion behavior is similar to bulk water for pore diameters between 2.0 nm – 2.7 nm (106). CMD simulations have also shown that $H_2O$ and ion mobility in $SiO_2$ slit nanopores strongly depend on the surface protonation state: $H_2O$ self-diffusion coefficients increase 5-fold when the fully de-protonated $SiO_2$ surface is compared to the fully protonated one (107).

## 3. Adsorption in nanopores

As discussed in sections 2.1 and 2.2, surfaces impose structure on solutes and $H_2O$ inside nanopores; because of this structure, the $\Delta G_{solv}$ for a nanoconfined species may be less negative than that for an unconfined counterpart (108; 109):

$$\Delta G_{solv} = z_i^2 e^2 / r_i \left( \frac{1}{\varepsilon_p} - \frac{1}{\varepsilon_b} \right) \text{ (Eq. 2)}$$

where $\Delta G_{solv}$ depends on ion charge $z_i$, the radius of the solvated ion $r_i$ (including $H_2Os$), the dielectric constants of fluid in the nanopore $\varepsilon_p$ and bulk solution $\varepsilon_b$, and the electronic charge constant $e$. When $\Delta G_{solv}$ becomes *less* negative, the Gibbs free energy of adsorption $\Delta G_{ads}$ may become *more* negative, because $\Delta G_{ads}$ is the sum of the electrostatic and chemical free energy

changes (favorable to adsorption) and the solvation free energy change (unfavorable to adsorption) (110):

$$\Delta G_{ads} = \Delta G_{coul} + \Delta G_{solv} + \Delta G_{chem} \quad \text{(Eq. 3)}$$

where $\Delta G_{coul}$ depends on the electrostatic interactions between the adsorbing species and the near-surface ions, $\Delta G_{solv}$ is the change in free energy of solvation for the adsorbing ion and the surface, and $\Delta G_{chem}$ is a binding energy that includes chemical bonding. More negative $\Delta G_{ads}$ shifts the adsorption equilibrium constant $K_{eq}$ toward products and favors the formation of products that require a desolvation step.

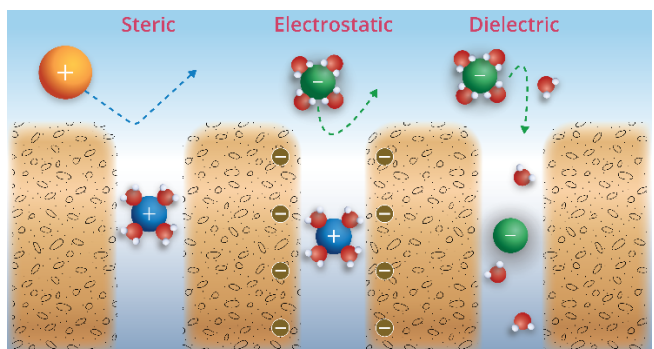

**Figure 7**. Chemical forces in membrane nanopores and their control on solute permeation. Modified from Ref. (5).

Studies that span separation science (nanoporous membranes), material science (carbon nanotubes, metal organic frameworks), and geoscience (oxide, zeolite, and clay mineral nanopores) show that nanoconfinement affects the thermodynamics (enthalpy and Gibbs free energy and corresponding $K_{eq}$ values), rates, and products of adsorption. In the systems reviewed, either outer-sphere or inner-sphere complexation reactions take place (Figure 2). In studies of ion diffusion in clay mineral interlayers and ion permeation through membrane nanopores [including polymers (e.g. polyamide), porous oxides ($SiO_2$, $Al_2O_3$, $TiO_2$) and zeolites (5; 111; 112)] nanoconfinement has a strong impact (19; 108; 113-115). As discussed in section

2.5, solute-surface and solute-water interactions in nanopores define effective diffusion/permeation rates. In membrane studies, several molecular forces are proposed to explain macroscopically-observed ion-specific diffusion rates and selective permeation: (i) steric rejection, or size-exclusion effects (112; 116), (ii) Charge-exclusion effects (117), and (iii) dielectric-based exclusion (112; 116) (Figure 7). The molecular origins of the dielectric-based exclusion mechanisms are poorly understood (112); one explanation is the change in energy barriers for solute desolvation under nanoconfinement (118). The energy barrier ($\Delta W_i$) depends on ion charge ($z_i$), hydrodynamic radius of solute $i$ ($a_s$, m), and the difference in the dielectric constants of the fluids in nanopores ($\varepsilon_p$, J$^{-1}$C$^2$m$^{-1}$) and bulk water ($\varepsilon_b$, J$^{-1}$C$^2$m$^{-1}$):

$$\Delta W_i = \frac{z_i^2 e^2}{8\pi\varepsilon_0 a_s}\left(\frac{1}{\varepsilon_p} - \frac{1}{\varepsilon_b}\right) \quad \text{(Eq. 4)}$$

where $e$ is the elementary charge and $\varepsilon_0$ is the permittivity of free space. As the pore diameter decreases, $\frac{1}{\varepsilon_p}$ and the energy barrier for ions entering the pore increases; this explains salt rejection in progressively smaller pores (118) and predicts that the energy barrier increases steeper for ions with larger valence. We argue below that, for charged species, the energy barrier for entering a nanopore may increase with decreasing pore diameter, but the free energy of adsorption $\Delta G_{ads}$ inside a nanopore is typically more favorable than on an unconfined surface.

## 3.1 Adsorption complexes

Several studies have found that inner-sphere cation adsorption is enhanced in nanopores. This enhancement is observed in both small (<1 nm) nanopores with diameters near the hydrated diameter of a cation (114; 115; 119-121), as well as in nanopores >1 nm (109; 122). Increased inner-sphere complexation was initially reported for zeolites with ~0.3-0.5 nm pores for $Na^+$, $Ca^{2+}$, $Ni^{2+}$, $Cu^{2+}$, and $Mn^{2+}$ based on batch experiments, flow microcalorimetry, NMR, and electron paramagnetic resonance spectroscopy measurements (114; 115; 119-121). These observations showed that in progressively smaller zeolite pores monovalent $Na^+$ cations with lower $\Delta G_{solv}$ are selected over the divalent $Ca^{2+}$ and $Ni^{2+}$ cations with more negative $\Delta G_{solv}$. Also, $Na^+$ and $Ca^{2+}$ adsorption as either outer-sphere or inner-sphere complexes is dictated by the zeolite pore diameter: in 0.74 nm pores, both cations adsorb as outer-sphere complexes, while in 0.51-0.56 nm pores $Na^+$ forms inner-sphere complexes, and $Ca^{2+}$ forms outer-sphere complexes (115). The nanopore inner-sphere enhancement (NISE) theory formulated based on these observations, states that weakly hydrated ions will be adsorbed preferentially in nanopores smaller than the hydrated ion (114; 115; 119). Thus, NISE theory predicts that the ratio (solvated ion diameter):(nanopore diameter) defines the strength of ion-surface interaction (119).

Therefore, to determine whether the primary descriptor of adsorption in nanopores is ion size, or $\Delta G_{solv}$, Ilgen et al. exploited the gradual changes in ion size and $\Delta G_{solv}$ for trivalent lanthanide ($Ln^{3+}$) cations. $Ln^{3+}$ adsorption was measured in templated $SiO_2$ nanopores with 4 and 7 nm pore diameters (109), the adsorption energetics (enthalpy $\Delta H_{ads}$) was quantified using flow microcalorimetry, and the local coordination environment of adsorbed $Ln^{3+}$ cations ($Nd^{3+}$, $Tb^{3+}$, and $Lu^{3+}$) on both confined and unconfined $SiO_2$ surfaces was determined with X-ray absorption fine structure (XAFS) spectroscopy (109). The $\Delta G_{solv}$ of examined $Ln^{3+}$ cations in bulk water range from -3280 to -3515 kJ $mol^{-1}$ (123). Ilgen et al. report that, as the nanopore size and hence the dielectric response of water decreases, the $\Delta G_{solv}$ values also decrease, promoting inner-sphere complex formation that requires the partial de-solvation of $Ln^{3+}$ and/or surface site prior to adsorption (109). Furthermore, more polymeric surface species form within $SiO_2$ nanopores, compared to unconfined surfaces (109; 122). This study shows that nanoconfinement effects are greater for $Ln^{3+}$ cations with less negative $\Delta G_{solv}$ ($Nd^{3+}$, $Eu^{3+}$, $Tb^{3+}$), than more negative $\Delta G_{solv}$ ($Tm^{3+}$, $Lu^{3+}$). The radii of examined solvated $Ln^{3+}$ ions range from 2.32 Å (9-coordinated $Nd^{3+}$) to 1.96 Å (8-coordinated $Lu^{3+}$) (123), and are significantly smaller than the $SiO_2$ nanopore diameters of 40 and 70 Å (4 and 7 nm). Therefore, $\Delta G_{solv}$ of adsorbing cations is the main descriptor for predicting nanoconfinement-driven changes in adsorption isotherms and surface complexation products (109).

Analogous studies of divalent $Cu^{2+}$ and $Fe^{2+}/Fe^{3+}$ adsorption in 4 and 7 nm $SiO_2$ nanopores (113; 122) combined experimental measurements and CMD simulations to show that $Cu^{2+}$ and $Fe^{2+}/Fe^{3+}$ cations adsorb at $SiO^-$ sites on amorphous $SiO_2$ as inner-sphere complexes (122; 124). The CMD simulations indicate that inner-sphere adsorption of $Cu^{2+}$ and $Fe^{2+}/Fe^{3+}$ increases with decreasing pore diameter (122; 124); and that $SiO_2$ surface charge exerts the primary control on $Fe^{2+}/Fe^{3+}$ adsorption, in agreement with the trends observed for unconfined surfaces (124).

In contrast, during $Zn^{2+}$ adsorption on controlled pore glass ($SiO_2$) with larger pores than in the studies above (~10 to 330 nm pore diameters), no enhancement in inner-sphere adsorption was observed (108). In all cases, $Zn^{2+}$ cations form inner-sphere mono-dentate complexes with $SiO_2$ substrates, characterized using XAFS. For confined and unconfined $SiO_2$ surfaces, the ratio of tetrahedral/octahedral $Zn^{2+}$ complexes increases with increasing surface coverage (108). Importantly, for $SiO_2$ nanopores with ~10 nm diameter, the change from predominantly octahedral to tetrahedral $Zn^{2+}$ complexes happens at a lower surface coverage, than on unconfined $SiO_2$ surfaces. Two mechanisms may explain the observed increase in the tetrahedral-to-octahedral ratio in smaller nanopores: the decrease in $\varDelta G_{solv}$, which favors adsorbed $Zn^{2+}$ with fewer $H_2Os$ in the 1$^{st}$ shell; and the potential change in surface chemistry of available surface Si-OH sites with decreasing pore diameter (e.g., charge, $SiO_2$ polymerization). Additionally, while $Zn^{2+}$ forms mono-dentate complexes (tetrahedral or octahedral), it promotes de-protonation of two surface sites: the Si-OH site where $Zn^{2+}$ adsorbs, and a neighboring site. This study reports no macroscopically observed shifts in the pH-dependent adsorption with change in $SiO_2$ pore diameters (108). However, Ilgen *et al.* and Knight *et al.* documented macroscopic pore-size dependent trends for $Cu^{2+}$ (113; 122), and for $Ln^{3+}$ cations with lower $\varDelta G_{solv}$ ($Nd^{3+}$, $Eu^{3+}$, $Tb^{3+}$) , but not for $Ln^{3+}$ with higher $\varDelta G_{solv}$ ($Tm^{3+}$, $Lu^{3+}$) (109).

Other experimental studies inferred surface complex geometry based on macroscopic trends by examining $Cd^{2+}$ adsorption on non-porous $SiO_2$ and $SiO_2$ with 3.8-6.4 nm pores (125), and U(VI) species [here U(VI) refers to all aqueous species] adsorption on non-porous α-$Al_2O_3$ (corundum) and $Al_2O_3$ with ~ 1.3 nm pores (126). The surface-area-normalized adsorption of $Cd^{2+}$ is higher for non-porous than porous $SiO_2$; however, when adsorption is normalized by Si-OH surface site density instead, adsorption becomes higher for $SiO_2$ nanopores; inner-sphere complex where $Cd^{2+}$ coordinates to two Si-OH sites (bi-dentate) was proposed (125). On porous $Al_2O_3$, U(VI) adsorption is pH-dependent, but independent of ionic strength, indicating inner-sphere complexation. However, on corundum, U(VI) adsorption depends on both pH and ionic strength, suggesting outer-sphere complexation (126). This observation is corroborated by sequential desorption experiments, that show that the majority of U(VI) is adsorbed irreversibly onto $Al_2O_3$ nanopores (only 5% can be desorbed), while 100% of U(VI) can be desorbed from corundum surfaces. The non-reversible U(VI) adsorption on $Al_2O_3$ nanopores was confirmed by other studies showing that U(VI) adsorbed to $Al_2O_3$ nanopores cannot be reduced to U(IV), but U(VI) adsorbed on corundum can be readily reduced to U(IV) (127).

These collective findings illustrate that the speciation of adsorbed cations depends on nanoconfinement; however, speciation is not easily anticipated since findings from one cation cannot be directly applied to other cations. This is due to the limited number of studies, and differences in structures and chemistries of nanopores and aqueous solutions; with more observations collected on comparable systems, general trends could be identified in the future. We found no studies investigating anion adsorption and their corresponding surface speciation in nanopores, which should be assessed in future studies.

## 3.2 Adsorption thermodynamics

Adsorption thermodynamics is frequently measured using either immersion calorimetry (when the sorbent is immersed into reactive solution) or flow microcalorimetry measurements (where a column is packed with solid substrate and heats are measured during ion-exchange reactions) (109; 115; 122). Flow microcalorimetry yields a cumulative heat signal, averaged between (de)solvation, physisorption, and chemisorption processes. In previous flow microcalorimetry studies of $Cu^{2+}$ and $Ln^{3+}$, cation adsorption was *endothermic* for unconfined surfaces and strongly *exothermic* in $SiO_2$ nanopores (19; 109). Ilgen *et al.* proposed that $\Delta H_{ads}$ shifts (i.e. adsorption produces more heat in progressively smaller pores), due to a combination of two processes. First, the lower average dielectric response of nanoconfined water (22; 24; 39) decreases $\Delta G_{solv}$, decreasing the cost of removing 1-2 $H_2O$ molecules from the solvation shell of a cation (109). Second, additional heat may be produced because adsorbing cations ($Cu^{2+}$, $Nd^{3+}$, $Tb^{3+}$, and $Lu^{3+}$) form more dimers in nanopores than on unconfined surfaces (19; 109).

Similarly, Wu and Navrotsky quantified $\Delta H_{ads}$ of small molecule interactions with $SiO_2$ nanopores using immersion calorimetry and found that $\Delta H_{ads}$ becomes more negative as pore diameter decreases and Si–OH surface densities increase. When porous $SiO_2$ is immersed in NaCl solution, $\Delta H_{ads}$ shifts were consistently observed for water, ethanol, triethylamine, sodium chloride, and sodium (bi)carbonate solutions (128). For zeolites, adsorption of $Na^+$ and $Ca^{2+}$ produces no heat in 0.74 nm pores, while in 0.51-0.56 nm pores $Na^+$ adsorption is *exothermic* (inner-sphere), and $Ca^{2+}$ adsorption produces no heat (outer-sphere complexation) (115).

Contrary to the observations of *exothermic* adsorption of $Na^+$ (128), $Cu^{2+}$ and $Ln^{3+}$ in $SiO_2$ nanopores (109; 122), $Cd^{2+}$ adsorption is *endothermic* (and entropy-driven) for both non-porous $SiO_2$ and $SiO_2$ with 3.8-6.4 nm pore diameters (125). $\Delta H_{ads}$ decreases with decreasing pore diameter and the partial de-solvation of $Cd^{2+}$ cations prior to adsorption contributes to the overall $\Delta H_{ads}$ (125). These findings highlight that surface geometry/curvature matters because of the resulting mean distances between the available Si–OH surface sites. The most positive $\Delta H_{ads}$ values (*endothermic*) were observed for non-porous $SiO_2$, where $Cd^{2+}$ coordinated to 2 Si–OH surface sites (125). Thus, while $\Delta H_{ads}$ depends on nanoconfinement, it cannot be reliably predicted as a function of pore chemistry and pore diameter.

### 3.3 Adsorption kinetics and equilibrium constants

Wang *et al.* showed that the kinetics of acid-base titration equilibration are much slower for porous $Al_2O_3$ with ~2 nm pores (~ 4-5 minutes) than for non-porous $Al_2O_3$ (~0.5 minute) (129). Contrary to the predicted increase in the energy barrier for entry into nanopores (section 3.1), U(VI) adsorption equilibrium was faster for mesoporous $Al_2O_3$ (1 hour) than for non-porous $Al_2O_3$ (20 hours), although both cases follow pseudo-second-order kinetics (126). $Cu^{2+}$ adsorption kinetics on mesoporous $SiO_2$ also proved to be pore-size-dependent (113). The adsorption of $Cu^{2+}$ follows pseudo-first-order kinetics, so the rate constants increase as $SiO_2$ pore diameter decreases: $0.048 \pm 0.003$ µmol m$^{-2}$ min$^{-1}$ for 7 nm pores, $0.059 \pm 0.003$ µmol m$^{-2}$ min$^{-1}$ for 6 nm pores, and $0.071 \pm 0.003$ µmol m$^{-2}$ min$^{-1}$ for 4 nm pores (113). This kinetic data may indicate that $Cu^{2+}$ diffusion increases as pore diameter decreases (113).

Ilgen *et al.* quantified the equilibrium constants for lanthanide adsorption on non-porous $SiO_2$ and $SiO_2$ nanopores (109). The surface-area-normalized equilibrium $K_{eq}$ for trivalent lanthanide cations vary linearly as a function of atomic mass and surface loading for unconfined $SiO_2$; however, in $SiO_2$ nanopores they deviate from linearity (Figure 8). Remarkably, the nanoconfinement effects are non-uniform across the $Ln^{3+}$ series: the lighter $Ln^{3+}$ show pore-size dependent behaviors. Nanoconfinement shifts the equilibrium adsorption constants more dramatically for the lighter lanthanides (with less negative $\Delta G_{solv}$) than the heavier ones (with more negative $\Delta G_{solv}$).

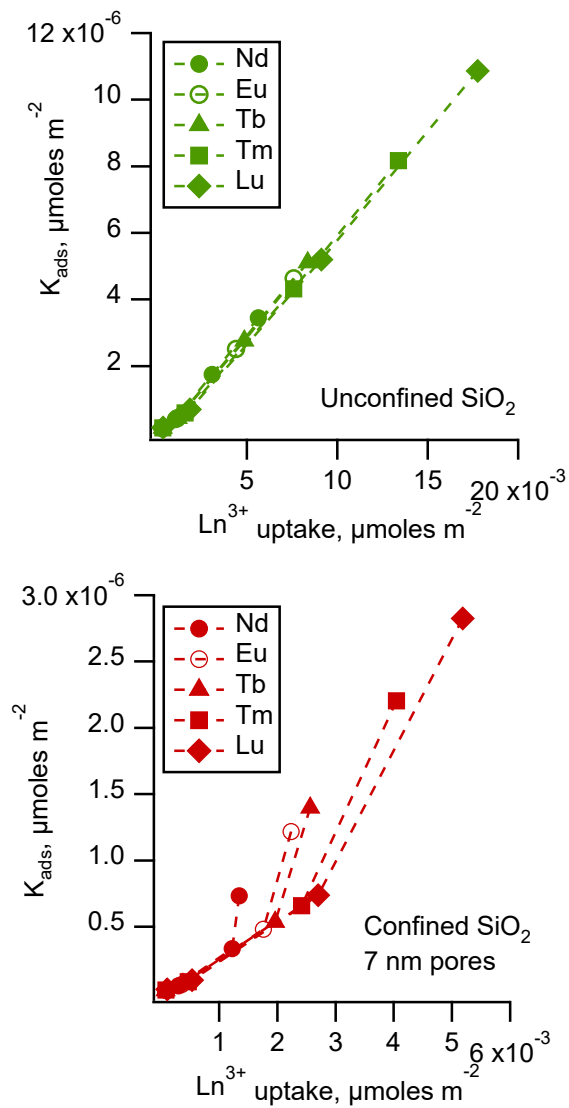

**Figure 8**. Surface-area-normalized equilibrium constants ($K_{eq}$) for lanthanide adsorption on unconfined $SiO_2$ (top panel) and confined $SiO_2$ surfaces with 7 nm pores (bottom panel). From Ref. (109). Images reprinted with permission from the Royal Chemical Society.

Wang *et al*. discovered that points of zero charge (PZC) were similar for 2-nm porous and non-porous γ-$Al_2O_3$; however, the surface charge density was ~2 times higher for porous $Al_2O_3$ than non-porous $Al_2O_3$ (129) and there were marked differences in the surface de-protonation constants *pKa* for porous $Al_2O_3$ (9.0 and 10.3) and non-porous $Al_2O_3$ (7.7 and 11.0) (129). Similar narrowing of the pH range (the difference between the two *pKa* values) under nanoconfinement was reported by Sun *et al.* (126), who also used acid-base titrations to measure PZC and *pKa* values for $Al_2O_3$ with ~ 1.3 nm pores (10.8 and 11.0), and non-porous corundum (8.9 and 10.9); the reported PZC values were 10.9 for porous $Al_2O_3$, and 9.7 for corundum.

The effect of nanoconfinement on the *pKa* of surface OH groups has been challenging to compute with first principles calculations. The deprotonation free energies can be computed using AIMD PMF simulations and related to the *pKa* of the surface OH groups (130). However, due to the computational cost, AIMD simulation cell sizes are generally limited to <2 nm thickness of water region; confinement effects await future quantification. However, experimental studies of nanoconfinement on stacks of 2-D transition metal carbide sheets at <1 nm spacings have been conducted using synthetic MXene materials terminated by an oxygen plane with hydroxylated basal surfaces. The outer oxygenated basal planes of MXenes can form heterostructures with other 2-D materials like graphene and titanium oxide (131; 132). The competition between proton diffusion and redox reactions on the hydroxylated surfaces were examined both theoretically and experimentally (133). Water structure, dynamics, and hydration structures of ions were predicted (134). However, the *pKa* values of surface groups have not been reported in these MXene systems and warrant future research.

## 4. Outlook: Challenges and Opportunities

The experimental and modeling studies of cation adsorption in nanopores indicate that the pathways, thermodynamics, and products of adsorption are dictated by (i) surface chemistry and topology; (ii) pore diameter and the corresponding structure of overlapping EDLs; and (iii) Gibbs free energy of ion solvation $\Delta G_{solv}$ (109; 114; 119). Current models struggle to predict the combined effects of nanoconfinement and surface chemistry, and how these effects manifest in macroscopic thermodynamics and diffusion rates in membrane and geological nanopores (63; 108; 109; 113; 122; 124; 135-139). The challenge for developing such descriptions is the dynamic and transient nature of the confining surfaces and the corresponding dynamic changes in $H_2O$ and solute speciation.

Future research should deepen our understanding of the dielectric responses and HB networks of nanoconfined $H_2O$: how they change with changing pore dimensions and molecular details of confining surfaces (net charge, charge distribution, presence of near-surface electrolytes). We also need to understand how the details of the dielectric response and HB network structures affect $H_2O$ reactivities (i.e. $H_2O$ self-dissociation) and the $\Delta G_{solv}$ of aqueous species, which will require thermodynamic integration methods to elucidate surface/interfacial potentials at these low dimensions (140). A complementary study might use PMF calculations to predict free energy changes as ions move from a bulk-like water into a connected nanoconfined region (141). It will also be crucial to model relevant system sizes to capture long-range (>10 nm) interactions that are currently only available to probe experimentally (14; 19; 27; 108; 109). To enable investigation of larger pores (~100 nm) at lower (e.g. micromolar) solute concentrations, system sizes for CMD models and simulation times must be significantly increased (142), possibly by leveraging recent advances in accelerated MD methods (143) and machine learned models of interatomic potentials (144; 145). Similarly, machine learning methods with quantum-level accuracy (146-148) would enable the prediction of chemical reactions under realistic nanoconfined conditions. Finally, dimensionality must be rigorously evaluated to reveal how the derivations of dielectric profiles and HB networks differ in slit pores (1-D) compared to cylindrical pores (2-D), and cages (3-D).

By developing a clear picture of nanoconfined $H_2O$ and solutes in 1-D, 2-D, and 3-D nanoconfined systems, we can advance our understanding of reactions in nanopores, including adsorption processes. Developing a predictive framework for pathways, products, and $K_{eq}$ values for adsorption in nanopores presents a major challenge and an opportunity. For example, there are no current studies on anion adsorption, which will be particularly relevant to radionuclide migration from underground nuclear waste repositories. New knowledge about adsorption in nanopores is critically needed to advance separation science, water treatment, environmental clean-up, and numerous other applications. Moreover, since adsorption complexes, thermodynamics, and adsorption affinity change with pore diameter, nanopores with tunable sizes and surface chemistries may enable the separation of critical elements from their aqueous mixtures (149; 150).

## Acknowledgements

The authors thank I. Sava-Gallis and R.T. Cygan for pre-submission review of this manuscript. This material is based upon work supported by the U.S. Department of Energy, Office of Science, Office of Basic Energy Sciences, Chemical Sciences, Geosciences, and Biosciences Division under Field Work Proposal Number 21-015452 at Sandia National Laboratories. Sandia National Laboratories is a multimission laboratory managed and operated by National Technology and Engineering Solutions of Sandia LLC, a wholly owned subsidiary of Honeywell International Inc., for the U.S. Department of Energy's National Nuclear Security Administration under contract DE-NA-0003525. This review paper describes objective technical results and analysis. Any subjective views or opinions that might be expressed in the paper do not necessarily represent the views of the U.S. Department of Energy or the United States Government.

# References

1. Wang Y. 2014. Nanogeochemistry: nanostructures, emergent properties and their control on geochemical reactions and mass transfers. *Chemical Geology* 378:1-23
2. Caro J. 2016. Hierarchy in inorganic membranes. *Chemical Society Reviews* 45:3468-78
3. Tournassat C, Steefel CI, Bourg IC, Bergaya F. 2015. *Natural and engineered clay barriers*. Elsevier
4. Kosinov N, Gascon J, Kapteijn F, Hensen EJ. 2016. Recent developments in zeolite membranes for gas separation. *Journal of Membrane Science* 499:65-79
5. Epsztein R, DuChanois RM, Ritt CL, Noy A, Elimelech M. 2020. Towards single-species selectivity of membranes with subnanometre pores. *Nature Nanotechnology* 15:426-36
6. Wu J, Li Z, Tan H, Du S, Liu T, et al. 2020. Highly Selective Separation of Rare Earth Elements by Zn-BTC Metal–Organic Framework/Nanoporous Graphene via In Situ Green Synthesis. *Analytical Chemistry* 93:1732-9
7. Na K, Somorjai GA. 2015. Hierarchically nanoporous zeolites and their heterogeneous catalysis: current status and future perspectives. *Catalysis Letters* 145:193-213
8. An Y, Tian Y, Ci L, Xiong S, Feng J, Qian Y. 2018. Micron-sized nanoporous antimony with tunable porosity for high-performance potassium-ion batteries. *ACS nano* 12:12932-40
9. Albrecht T. 2019. Single-molecule analysis with solid-state nanopores. *Annual Review of Analytical Chemistry* 12:371-87
10. Baxter JB, Richter C, Schmuttenmaer CA. 2014. Ultrafast carrier dynamics in nanostructures for solar fuels. *Annual review of physical chemistry* 65:423-47
11. Zhou K, Perry JM, Jacobson SC. 2011. Transport and sensing in nanofluidic devices. *Annual Review of Analytical Chemistry* 4:321-41
12. Senanayake HS, Greathouse JA, Ilgen AG, Thompson WH. 2021. Simulations of the IR and Raman spectra of water confined in amorphous silica slit pores. *The Journal of Chemical Physics* 154:104503
13. Senapati S, Chandra A. 2001. Dielectric constant of water confined in a nanocavity. *The Journal of Physical Chemistry B* 105:5106-9
14. Tsukahara T, Hibara A, Ikeda Y, Kitamori T. 2007. NMR study of water molecules confined in extended nanospaces. *Angewandte Chemie* 119:1199-202
15. Zhang MS, Mao H, Jin ZH. 2021. Molecular dynamic study on structural and dynamic properties of water, counter-ions and polyethylene glycols in Na-montmorillonite interlayers. *Applied Surface Science* 536:147700
16. Baum M, Rieutord F, Juranyi F, Rey C, Rébiscoul D. 2019. Dynamical and structural properties of water in silica nanoconfinement: impact of pore size, ion nature, and electrolyte concentration. *Langmuir* 35:10780-94
17. Musat R, Renault JP, Candelaresi M, Palmer DJ, Le Caër S, et al. 2008. Finite size effects on hydrogen bonds in confined water. *Angewandte Chemie International Edition* 47:8033-5
18. Brubach J-B, Mermet A, Filabozzi A, Gerschel A, Lairez D, et al. 2001. Dependence of water dynamics upon confinement size. *The Journal of Physical Chemistry B* 105:430-5
19. Knight AW, Ilani-Kashkouli P, Harvey JA, Greathouse JA, Ho TA, et al. 2020. Interfacial reactions of Cu (ii) adsorption and hydrolysis driven by nano-scale confinement. *Environmental Science: Nano* 7:68-80

20. Knight AW, Kalugin NG, Coker E, Ilgen AG. 2019. Water properties under nano-scale confinement. *Scientific reports* 9:1-12
21. Israelachvili JN. 2011. *Intermolecular and surface forces*. Academic press
22. Fumagalli L, Esfandiar A, Fabregas R, Hu S, Ares P, et al. 2018. Anomalously low dielectric constant of confined water. *Science* 360:1339-42
23. Brovchenko I, Oleinikova A. 2008. *Interfacial and confined water*. Elsevier
24. Morikawa K, Kazoe Y, Mawatari K, Tsukahara T, Kitamori T. 2015. Dielectric constant of liquids confined in the extended nanospace measured by a streaming potential method. *Analytical chemistry* 87:1475-9
25. Debye P. 1923. The theory of electrolytes. I. *Z. Phys.* 24:305-24
26. Levinger NE. 2002. Water in confinement. *Science* 298:1722-3
27. Takei T, Mukasa K, Kofuji M, Fuji M, Watanabe T, et al. 2000. Changes in density and surface tension of water in silica pores. *Colloid and Polymer Science* 278:475-80
28. Le Caër S, Pin S, Esnouf S, Raffy Q, Renault JP, et al. 2011. A trapped water network in nanoporous material: the role of interfaces. *Physical Chemistry Chemical Physics* 13:17658-66
29. Marti J, Nagy G, Guardia E, Gordillo M. 2006. Molecular dynamics simulation of liquid water confined inside graphite channels: dielectric and dynamical properties. *The Journal of Physical Chemistry B* 110:23987-94
30. Koga K, Gao G, Tanaka H, Zeng XC. 2001. Formation of ordered ice nanotubes inside carbon nanotubes. *Nature* 412:802-5
31. Striolo A, Chialvo A, Gubbins K, Cummings P. 2005. Water in carbon nanotubes: Adsorption isotherms and thermodynamic properties from molecular simulation. *The Journal of chemical physics* 122:234712
32. Hirunsit P, Balbuena PB. 2007. Effects of confinement on water structure and dynamics: a molecular simulation study. *The Journal of Physical Chemistry C* 111:1709-15
33. Cicero G, Grossman JC, Schwegler E, Gygi F, Galli G. 2008. Water confined in nanotubes and between graphene sheets: A first principle study. *Journal of the American Chemical Society* 130:1871-8
34. Rasaiah JC, Garde S, Hummer G. 2008. Water in nonpolar confinement: From nanotubes to proteins and beyond. *Annu. Rev. Phys. Chem.* 59:713-40
35. Alexiadis A, Kassinos S. 2008. Molecular simulation of water in carbon nanotubes. *Chemical reviews* 108:5014-34
36. Chakraborty S, Kumar H, Dasgupta C, Maiti PK. 2017. Confined water: structure, dynamics, and thermodynamics. *Accounts of chemical research* 50:2139-46
37. Ruiz Pestana L, Felberg LE, Head-Gordon T. 2018. Coexistence of multilayered phases of confined water: the importance of flexible confining surfaces. *ACS nano* 12:448-54
38. Zaragoza A, González MA, Joly L, López-Montero I, Canales M, et al. 2019. Molecular dynamics study of nanoconfined TIP4P/2005 water: how confinement and temperature affect diffusion and viscosity. *Physical Chemistry Chemical Physics* 21:13653-67
39. Motevaselian MH, Aluru NR. 2020. Universal Reduction in Dielectric Response of Confined Fluids. *ACS nano* 14:12761-70
40. Ortiz-Young D, Chiu H-C, Kim S, Voïtchovsky K, Riedo E. 2013. The interplay between apparent viscosity and wettability in nanoconfined water. *Nature communications* 4:1-6
41. Sansom MS, Biggin PC. 2001. Water at the nanoscale. *Nature* 414:157-9

42. Mattia D, Calabrò F. 2012. Explaining high flow rate of water in carbon nanotubes via solid–liquid molecular interactions. *Microfluidics and nanofluidics* 13:125-30
43. Parsons DF, Boström M, Nostro PL, Ninham BW. 2011. Hofmeister effects: interplay of hydration, nonelectrostatic potentials, and ion size. *Physical Chemistry Chemical Physics* 13:12352-67
44. Acuña SM, Toledo PG. 2011. Nanoscale repulsive forces between mica and silica surfaces in aqueous solutions. *Journal of colloid and interface science* 361:397-9
45. Hunter RJ. 2001. *Foundations of colloid science*. Oxford university press
46. Li D, Chun J, Xiao D, Zhou W, Cai H, et al. 2017. Trends in mica–mica adhesion reflect the influence of molecular details on long-range dispersion forces underlying aggregation and coalignment. *Proceedings of the National Academy of Sciences* 114:7537-42
47. Prakash A, Pfaendtner J, Chun J, Mundy CJ. 2017. Quantifying the molecular-scale aqueous response to the mica surface. *The Journal of Physical Chemistry C* 121:18496-504
48. Zachariah Z, Espinosa-Marzal RM, Heuberger MP. 2017. Ion specific hydration in nano-confined electrical double layers. *Journal of colloid and interface science* 506:263-70
49. van Lin SR, Grotz KK, Siretanu I, Schwierz N, Mugele F. 2019. Ion-specific and ph-dependent hydration of mica–electrolyte interfaces. *Langmuir* 35:5737-45
50. Gilbert B, Comolli LR, Tinnacher RM, Kunz M, Banfield JF. 2015. Formation and restacking of disordered smectite osmotic hydrates. *Clays and Clay Minerals* 63:432-42
51. Ho TA, Criscenti LJ. 2021. Molecular-level understanding of gibbsite particle aggregation in water. *Journal of Colloid and Interface Science* 600:310-7
52. Gonçalvès J, Rousseau-Gueutin P, Revil A. 2007. Introducing interacting diffuse layers in TLM calculations: A reappraisal of the influence of the pore size on the swelling pressure and the osmotic efficiency of compacted bentonites. *Journal of colloid and interface science* 316:92-9
53. Gonçalvès J, Rousseau-Gueutin P. 2008. Molecular-scale model for the mass density of electrolyte solutions bound by clay surfaces: Application to bentonites. *Journal of colloid and interface science* 320:590-8
54. Shen X, Bourg IC. 2021. Molecular dynamics simulations of the colloidal interaction between smectite clay nanoparticles in liquid water. *Journal of Colloid and Interface Science* 584:610-21
55. Dragulet F, Goyal A, Ioannidou K, Pellenq RJ-M, Del Gado E. 2022. Ion specificity of confined ion-water structuring and nanoscale surface forces in clays. *arXiv preprint arXiv:2204.01631*
56. Pezzotti S, Serva A, Sebastiani F, Brigiano FS, Galimberti DR, et al. 2021. Molecular Fingerprints of Hydrophobicity at Aqueous Interfaces from Theory and Vibrational Spectroscopies. *The Journal of Physical Chemistry Letters* 12:3827-36
57. Rovere M, Ricci M, Vellati D, Bruni F. 1998. A molecular dynamics simulation of water confined in a cylindrical $SiO_2$ pore. *The Journal of chemical physics* 108:9859-67
58. Rieth AJ, Hunter KM, Dincă M, Paesani F. 2019. Hydrogen bonding structure of confined water templated by a metal-organic framework with open metal sites. *Nature communications* 10:1-7
59. Hartnig C, Witschel W, Spohr E, Gallo P, Ricci MA, Rovere M. 2000. Modifications of the hydrogen bond network of liquid water in a cylindrical $SiO2$ pore. *Journal of Molecular Liquids* 85:127-37

60. Gordillo M, Martı J. 2000. Hydrogen bond structure of liquid water confined in nanotubes. *Chemical Physics Letters* 329:341-5
61. Medders GR, Paesani F. 2014. Water dynamics in metal–organic frameworks: Effects of heterogeneous confinement predicted by computational spectroscopy. *The Journal of Physical Chemistry Letters* 5:2897-902
62. Hare D, Sorensen C. 1992. Interoscillator coupling effects on the OH stretching band of liquid water. *The Journal of chemical physics* 96:13-22
63. Knight AW, Kalugin NG, Coker E, Ilgen AG. 2019. Water properties under nano-scale confinement. *Scientific reports* 9:8246
64. Malfait B, Moréac A, Jani A, Lefort R, Huber P, et al. 2022. Structure of Water at Hydrophilic and Hydrophobic Interfaces: Raman Spectroscopy of Water Confined in Periodic Mesoporous (Organo) Silicas. *The Journal of Physical Chemistry C* 126:3520-31
65. Rother G, Gautam S, Liu T, Cole DR, Busch A, Stack AG. 2022. Molecular Structure of Adsorbed Water Phases in Silica Nanopores. *The Journal of Physical Chemistry C* 126:2885-95
66. Muñoz-Santiburcio D, Marx D. 2021. Confinement-controlled aqueous chemistry within nanometric slit pores: Focus review. *Chemical Reviews* 121:6293-320
67. Teschke O, Ceotto G, De Souza E. 2000. Interfacial aqueous solutions dielectric constant measurements using atomic force microscopy. *Chemical Physics Letters* 326:328-34
68. Matyushov DV. 2021. Dielectric susceptibility of water in the interface. *The Journal of Physical Chemistry B* 125:8282-93
69. Loche P, Ayaz C, Wolde-Kidan A, Schlaich A, Netz RR. 2020. Universal and nonuniversal aspects of electrostatics in aqueous nanoconfinement. *The Journal of Physical Chemistry B* 124:4365-71
70. Ballenegger V, Hansen J-P. 2005. Dielectric permittivity profiles of confined polar fluids. *The Journal of chemical physics* 122:114711
71. Jalali H, Ghorbanfekr H, Hamid I, Neek-Amal M, Rashidi R, Peeters F. 2020. Out-of-plane permittivity of confined water. *Physical Review E* 102:022803
72. Jeanmairet G, Rotenberg B, Borgis D, Salanne M. 2019. Study of a water-graphene capacitor with molecular density functional theory. *The Journal of Chemical Physics* 151:124111
73. Jorn R, Kumar R, Abraham DP, Voth GA. 2013. Atomistic modeling of the electrode–electrolyte interface in Li-ion energy storage systems: electrolyte structuring. *The Journal of Physical Chemistry C* 117:3747-61
74. Hu B, Zhu H. 2021. Anomalous dielectric behaviors of electrolyte solutions confined in graphene oxide nanochannels. *Scientific Reports* 11:1-11
75. Jalali H, Lotfi E, Boya R, Neek-Amal M. 2021. Abnormal Dielectric Constant of Nanoconfined Water between Graphene Layers in the Presence of Salt. *The Journal of Physical Chemistry B* 125:1604-10
76. Azam MS, Weeraman CN, Gibbs-Davis JM. 2012. Specific cation effects on the bimodal acid–base behavior of the silica/water interface. *The journal of physical chemistry letters* 3:1269-74
77. Pfeiffer-Laplaud M, Gaigeot M-P, Sulpizi M. 2016. p K a at Quartz/Electrolyte Interfaces. *The Journal of Physical Chemistry Letters* 7:3229-34

78. Corti HR, Appignanesi GA, Barbosa MC, Bordin JR, Calero C, et al. 2021. Structure and dynamics of nanoconfined water and aqueous solutions. *The European Physical Journal E* 44:1-50
79. Muñoz-Santiburcio D, Wittekindt C, Marx D. 2013. Nanoconfinement effects on hydrated excess protons in layered materials. *Nature communications* 4
80. Di Pino S, Sirkin YAP, Morzan UN, Sánchez VM, Hassanali A, Scherlis DA. 2021. Water self-dissociation under the microscope: the Kw in confinement. *arXiv preprint arXiv:2104.12513*
81. Hashikawa Y, Hasegawa S, Murata Y. 2018. A single but hydrogen-bonded water molecule confined in an anisotropic subnanospace. *Chemical Communications* 54:13686-9
82. Liu X, Lu X, Wang R, Meijer EJ, Zhou H. 2011. Acidities of confined water in interlayer space of clay minerals. *Geochimica et Cosmochimica Acta* 75:4978-86
83. Sirkin YAP, Hassanali A, Scherlis DA. 2018. One-dimensional confinement inhibits water dissociation in carbon nanotubes. *The Journal of Physical Chemistry Letters* 9:5029-33
84. Grifoni E, Piccini G, Lercher JA, Glezakou V-A, Rousseau R, Parrinello M. 2021. Confinement effects and acid strength in zeolites. *Nature communications* 12:1-9
85. Le Caër S, Pignié M-C, Berrod Q, Grzimek V, Russina M, et al. 2021. Dynamics in hydrated inorganic nanotubes studied by neutron scattering: towards nanoreactors in water. *Nanoscale Advances* 3:789-99
86. Marry V, Dubois E, Malikova N, Breu J, Haussler W. 2013. Anisotropy of Water Dynamics in Clays: Insights from Molecular Simulations for Experimental QENS Analysis. *The Journal of Physical Chemistry C* 117:15106-15
87. Michot LJ, Ferrage E, Delville, Jimenez-Ruiz M. 2016. Influence of layer charge, hydration state and cation nature on the collective dynamics of interlayer water in synthetic swelling clay minerals. *Applied Clay Science* 119:375-84
88. Asaad A, Hubert F, Ferrage E, Dabat T, Paineau E, et al. 2021. Role of interlayer porosity and particle organization in the diffusion of water in swelling clays. *Applied Clay Science* 207:106089
89. Porion P, Warmont F, Faugere AM, Rollet AL, Dubois E, et al. 2015. Cs-133 Nuclear Magnetic Resonance Relaxometry as a Probe of the Mobility of Cesium Cations Confined within Dense Clay Sediments. *Journal of Physical Chemistry C* 119:15360-72
90. Nanda R, Bowers GM, Loganathan N, Burton SD, Kirkpatrick RJ. 2019. Temperature dependent structure and dynamics in smectite interlayers: Na-23 MAS NMR spectroscopy of Na-hectorite. *Rsc Advances* 9:12755-65
91. Holmboe M, Bourg IC. 2014. Molecular Dynamics Simulations of Water and Sodium Diffusion in Smectite Interlayer Nanopores as a Function of Pore Size and Temperature. *Journal of Physical Chemistry C* 118:1001-13
92. Collin M, Gin S, Dazas B, Mahadevan T, Du JC, Bourg IC. 2018. Molecular Dynamics Simulations of Water Structure and Diffusion in a 1 nm Diameter Silica Nanopore as a Function of Surface Charge and Alkali Metal Counterion Identity. *Journal of Physical Chemistry C* 122:17764-76
93. Greathouse JA, Hart DB, Bowers GM, Kirkpatrick RJ, Cygan RT. 2015. Molecular Simulation of Structure and Diffusion at Smectite–Water Interfaces: Using Expanded Clay Interlayers as Model Nanopores. *Journal of Physical Chemistry C* 119:17126-36

94. Simonnin P, Marry V, Noetinger B, Nieto-Draghi C, Rotenberg B. 2018. Mineral- and Ion-Specific Effects at Clay-Water Interfaces: Structure, Diffusion, and Hydrodynamics. *Journal of Physical Chemistry C* 122:18484-92
95. Botan A, Rotenberg B, Marry V, Turq P, Noetinger B. 2011. Hydrodynamics in Clay Nanopores. *The Journal of Physical Chemistry C* 115:16109-15
96. Martins DMS, Molinari M, Goncalves MA, Mirao JP, Parker SC. 2014. Toward Modeling Clay Mineral Nanoparticles: The Edge Surfaces of Pyrophyllite and Their Interaction with Water. *Journal of Physical Chemistry C* 118:27308-17
97. Greathouse JA, Cygan RT, Fredrich JT, Jerauld GR. 2016. Molecular Dynamics Simulation of Diffusion and Electrical Conductivity in Montmorillonite Interlayers. *The Journal of Physical Chemistry C* 120:1640-9
98. Sánchez FtGl, Gimmi T, Jurányi F, Loon LV, Diamond LW. 2009. Linking the Diffusion of Water in Compacted Clays at Two Different Time Scales: Tracer Through-Diffusion and Quasielastic Neutron Scattering. *Environmental Science & Technology* 43:3487-93
99. Salles F, Douillard JM, Bildstein O, El Ghazi S, Prelot B, et al. 2015. Diffusion of Interlayer Cations in Swelling Clays as a Function of Water Content: Case of Montmorillonites Saturated with Alkali Cations. *Journal of Physical Chemistry C* 119:10370-8
100. Tinnacher RM, Holmboe M, Tournassat C, Bourg IC, Davis JA. 2016. Ion adsorption and diffusion in smectite: Molecular, pore, and continuum scale views. *Geochimica Et Cosmochimica Acta* 177:130-49
101. Underwood TR, Bourg IC. 2020. Large-Scale Molecular Dynamics Simulation of the Dehydration of a Suspension of Smectite Clay Nanoparticles. *Journal of Physical Chemistry C* 124:3702-14
102. Scalfi L, Fraux G, Boutin A, Coudert FX. 2018. Structure and Dynamics of Water Confined in Imogolite Nanotubes. *Langmuir* 34:6748-56
103. González RI, Rojas-Nunez J, Valencia FJ, Munoz F, Baltazar SE, et al. 2020. Imogolite in water: Simulating the effects of nanotube curvature on structure and dynamics. *Applied Clay Science* 191:105582
104. Fernandez-Martinez A, Tao JH, Wallace AF, Bourg IC, Johnson MR, et al. 2020. Curvature-induced hydrophobicity at imogolite-water interfaces. *Environmental Science-Nano* 7:2759-72
105. Faraone A, Liu K-H, Mou C-Y, Zhang Y, Chen S-H. 2009. Single particle dynamics of water confined in a hydrophobically modified MCM-41-S nanoporous matrix. *The Journal of Chemical Physics* 130:134512
106. Briman IM, Rébiscoul D, Diat O, Zanotti J-M, Jollivet P, et al. 2012. Impact of Pore Size and Pore Surface Composition on the Dynamics of Confined Water in Highly Ordered Porous Silica. *The Journal of Physical Chemistry C* 116:7021-8
107. Ho TA, Argyris D, Cole DR, Striolo A. 2012. Aqueous NaCl and CsCl solutions confined in crystalline slit-shaped silica nanopores of varying degree of protonation. *Langmuir* 28:1256-66
108. Nelson J, Bargar JR, Wasylenki L, Brown Jr GE, Maher K. 2018. Effects of nano-confinement on Zn (II) adsorption to nanoporous silica. *Geochimica et Cosmochimica Acta* 240:80-97

109. Ilgen AG, Kabengi N, Leung K, Ilani-Kashkouli P, Knight AW, Loera L. 2021. Defining silica–water interfacial chemistry under nanoconfinement using lanthanides. *Environmental Science: Nano* 8:432-43
110. James RO, Healy TW. 1972. Adsorption of hydrolyzable metal ions at the oxide—water interface. III. A thermodynamic model of adsorption. *Journal of Colloid and Interface Science* 40:65-81
111. Barry E, Burns R, Chen W, De Hoe GX, De Oca JMM, et al. 2021. Advanced Materials for Energy-Water Systems: The Central Role of Water/Solid Interfaces in Adsorption, Reactivity, and Transport. *Chemical Reviews* 121:9450-501
112. Shefer I, Peer-Haim O, Leifman O, Epsztein R. 2021. Enthalpic and Entropic Selectivity of Water and Small Ions in Polyamide Membranes. *Environmental Science & Technology*
113. Knight AW, Tigges A, Ilgen A. 2018. Adsorption of Copper on Mesoporous Silica: The Effect of Nano-scale Confinement. *Geochemical Transactions*
114. Ferreira D, Schulthess C. 2011. The nanopore inner sphere enhancement effect on cation adsorption: Sodium, potassium, and calcium. *Soil Science Society of America Journal* 75:389-96
115. Ferreira D, Schulthess C, Kabengi N. 2013. Calorimetric evidence in support of the nanopore inner sphere enhancement theory on cation adsorption. *Soil Science Society of America Journal* 77:94-9
116. Mohammad AW, Teow Y, Ang W, Chung Y, Oatley-Radcliffe D, Hilal N. 2015. Nanofiltration membranes review: Recent advances and future prospects. *Desalination* 356:226-54
117. Sujanani R, Landsman MR, Jiao S, Moon JD, Shell MS, et al. 2020. Designing solute-tailored selectivity in membranes: perspectives for water reuse and resource recovery. *ACS Macro Letters* 9:1709-17
118. Bowen WR, Welfoot JS. 2002. Modelling the performance of membrane nanofiltration—critical assessment and model development. *Chemical engineering science* 57:1121-37
119. Schulthess C, Taylor R, Ferreira D. 2011. The nanopore inner sphere enhancement effect on cation adsorption: Sodium and nickel. *Soil Science Society of America Journal* 75:378-88
120. Ferreira DR, Schulthess CP, Giotto MV. 2012. An investigation of strong sodium retention mechanisms in nanopore environments using nuclear magnetic resonance spectroscopy. *Environmental science & technology* 46:300-6
121. Ferreira DR, Schulthess CP, Amonette JE, Walter ED. 2012. An electron paramagnetic resonance spectroscopy investigation of the retention mechanisms of Mn and Cu in the nanopore channels of three zeolite minerals. *Clays and Clay Minerals* 60:588-98
122. Knight AW, Ilani-Kashkouli P, Harvey JA, Greathouse JA, Ho TA, et al. 2020. Interfacial reactions of Cu (ii) adsorption and hydrolysis driven by nano-scale confinement. *Environmental Science: Nano*
123. D'Angelo P, Spezia R. 2012. Hydration of lanthanoids (III) and actinoids (III): an experimental/theoretical saga. *Chemistry–A European Journal* 18:11162-78
124. Greathouse JA, Duncan TJ, Ilgen AG, Harvey JA, Criscenti LJ, Knight AW. 2021. Effects of nanoconfinement and surface charge on iron adsorption on mesoporous silica. *Environmental Science: Nano*
125. Prelot B, Lantenois S, Chorro C, Charbonnel M-C, Zajac J, Douillard JM. 2011. Effect of nanoscale pore space confinement on cadmium adsorption from aqueous solution onto ordered mesoporous silica: a combined adsorption and flow calorimetry study. *The Journal of Physical Chemistry C* 115:19686-95
126. Sun Y, Yang S, Sheng G, Guo Z, Tan X, et al. 2011. Comparison of U (VI) removal from contaminated groundwater by nanoporous alumina and non-nanoporous alumina. *Separation and purification technology* 83:196-203
127. Jung HB, Boyanov MI, Konishi H, Sun Y, Mishra B, et al. 2012. Redox behavior of uranium at the nanoporous aluminum oxide-water interface: Implications for uranium remediation. *Environmental science & technology* 46:7301-9
128. Wu D, Navrotsky A. 2013. Small molecule–Silica interactions in porous silica structures. *Geochimica et Cosmochimica Acta* 109:38-50
129. Wang Y, Bryan C, Xu H, Pohl P, Yang Y, Brinker CJ. 2002. Interface chemistry of nanostructured materials: Ion adsorption on mesoporous alumina. *Journal of colloid and interface science* 254:23-30
130. Leung K. 2021. First-principles Molecular Dynamics maps out complete mineral surface acidity landscape. *American Mineralogist: Journal of Earth and Planetary Materials* 106:1705-6
131. Xu L, Jiang D-e. 2021. Proton dynamics in water confined at the interface of the graphene–MXene heterostructure. *The Journal of Chemical Physics* 155:234707
132. Ganeshan K, Shin YK, Osti NC, Sun Y, Prenger K, et al. 2020. Structure and dynamics of aqueous electrolytes confined in 2D-TiO2/Ti3C2T2 MXene heterostructures. *ACS Applied Materials & Interfaces* 12:58378-89
133. Kobayashi T, Sun Y, Prenger K, Jiang D-e, Naguib M, Pruski M. 2020. Nature of Terminating Hydroxyl Groups and Intercalating Water in Ti3C2T x MXenes: A Study by 1H Solid-State NMR and DFT Calculations. *The Journal of Physical Chemistry C* 124:13649-55
134. Zhan C, Sun Y, Aydin F, Wang YM, Pham TA. 2021. Confinement effects on the solvation structure of solvated alkaline metal cations in a single-digit 1T-MoS2 nanochannel: A first-principles study. *The Journal of Chemical Physics* 154:164706
135. Rodriguez J, Elola MD, Laria D. 2009. Polar mixtures under nanoconfinement. *The Journal of Physical Chemistry B* 113:12744-9
136. Patra S, Pandey AK, Sarkar SK, Goswami A. 2014. Wonderful nanoconfinement effect on redox reaction equilibrium. *RSC Advances* 4:33366-9
137. Li L, Kohler F, Røyne A, Dysthe DK. 2017. Growth of Calcite in Confinement. *Crystals* 7:361
138. Jiang Q, Ward MD. 2014. Crystallization under nanoscale confinement. *Chemical Society Reviews* 43:2066-79
139. Argyris D, Cole DR, Striolo A. 2010. Ion-specific effects under confinement: the role of interfacial water. *Acs Nano* 4:2035-42
140. Wilson MA, Pohorille A, Pratt LR. 1988. Surface potential of the water liquid–vapor interface. *The Journal of chemical physics* 88:3281-5
141. Zhou K, Xu Z. 2020. Ion permeability and selectivity in composite nanochannels: engineering through the end effects. *The Journal of Physical Chemistry C* 124:4890-8
142. Alvarez F, Arbe A, Colmenero J. 2021. Unraveling the coherent dynamic structure factor of liquid water at the mesoscale by molecular dynamics simulations. *J. Chem. Phys.* 155:244509

143. Yang LJ, Liu CW, Shao Q, Zhang J, Gao YQ. 2015. From Thermodynamics to Kinetics: Enhanced Sampling of Rare Events. *Accounts of Chemical Research* 48:947-55
144. Jinnouchi R, Miwa K, Karsai F, Kresse G, Asahi R. 2020. On-the-Fly Active Learning of Interatomic Potentials for Large-Scale Atomistic Simulations. *Journal of Physical Chemistry Letters* 11:6946-55
145. Kang P-L, Shang C, Liu Z-P. 2020. Large-Scale Atomic Simulation via Machine Learning Potentials Constructed by Global Potential Energy Surface Exploration. *Accounts of Chemical Research* 53:2119-29
146. Keith JA, Vassilev-Galindo V, Cheng BQ, Chmiela S, Gastegger M, et al. 2021. Combining Machine Learning and Computational Chemistry for Predictive Insights Into Chemical Systems. *Chemical Reviews* 121:9816-72
147. Smith JS, Isayev O, Roitberg AE. 2017. ANI-1: an extensible neural network potential with DFT accuracy at force field computational cost. *Chemical Science* 8:3192-203
148. Noé F, Tkatchenko A, Müller K-R, Clementi C. 2020. Machine Learning for Molecular Simulation. *Annual Review of Physical Chemistry* 71:361-90
149. Ilgen A. 2020.
150. Lee Y-R, Yu K, Ravi S, Ahn W-S. 2018. Selective adsorption of rare earth elements over functionalized Cr-MIL-101. *ACS applied materials & interfaces* 10:23918-27